# Lorentz-covariant quantum transport and the origin of dark energy

**Arne Bergstrom**

Paper          pages    1 – 8

FAQ            pages    9 – 13

# Lorentz-covariant quantum transport and the origin of dark energy

**Arne Bergstrom**

B&E Scientific Ltd, Seaford BN25 4PA, United Kingdom

E-mail: arne.bergstrom@physics.org

**Abstract**
A possible explanation for the enigma of dark energy, responsible for about 76 % of the mass-energy of the universe, is obtained by requiring only that the rigorous continuity equation (the Boltzmann transport equation) for quanta propagating through space should have the form of a Lorentz-covariant and dispersion-free wave equation. This requirement implies (*i*) properties of space-time which an observer would describe as uniform expansion in agreement with Hubble's law, and (*ii*) that the quantum transport behaves like in a multiplicative medium with multiplication factor $v = 2$. This inherent, essentially explosive multiplicity of vacuum, thus caused by the requirement of Lorentz-covariance, is suggested as a potential origin of dark energy. In addition, it is shown (*iii*) that this requirement of Lorentz-covariant quantum transport leads to an apparent accelerated expansion of the universe.

(The figures in this article are in colour only in the electronic version.)

## 1. Introduction

Physics entered the new millennium with the enigma that the universe had just been found to be subjected to an accelerated expansion (Perlmutter, *et al.* [1], Riess, *et al.* [2]), as if caused by some dark energy. This added to the mystery inherent in the earlier discoveries by Fritz Zwicky [3], and later Very Rubin [4], of large amounts of dark matter around galaxies, and which discoveries had then finally begun to be taken seriously.

Thus, surprisingly, it dawned that most of the mass in the universe must be in the form of something hitherto unknown. The mass-energy of the universe is now considered to consist of about 76 % of dark energy and 20 % of dark matter [5], about the nature and origin of both there still seems to be no clear consensus [6] despite being a field of intense present interest [7−9].

At the present, the approach that seems to be most favoured to describe dark energy is a revival (one more time) of Einstein's cosmological constant Λ. The cosmological constant, it should be remembered, was originally introduced by Einstein as a "fudge factor" to permit steady-state solutions to his gravitational equations (which it actually didn't − since they were unstable − and which also turned out to be unnecessary anyway after Hubble had discovered that the universe indeed expands).

However, invoking the cosmological constant again might at least this time perhaps be supported theoretically by the fact pointed out by Zel'dovich [10] that the cosmological constant is mathematically equivalent to the stress-energy of vacuum, which in quantum field theory is filled with virtual particles. Unfortunately, however, calculations along these lines give estimates that are at least 60 orders of magnitude wrong [5]. The physical mechanism behind the cosmic acceleration thus still remains a deep mystery.

Recent results from the Seven-Year Wilkinson Microwave Anisotropy Probe (WMAP) and other experiments indicate that on cosmological scales the universe is flat, *i.e* has a density parameter $\Omega_{tot} = 1$ (corresponding to a cosmological constant $\Lambda = -1$), at least to within a very narrow error margin ($\Omega_{tot} = 1.0023 \pm 0.0056$) [11]. However, it should be remarked that on 'smaller' scales – up to the size of galaxies – gravitational effects are still important and dominate over the effects studied in this paper. Matter can then clump together under the influence of gravity, and these clumps – *e.g.*, galaxies – will then not expand individually [12] even though they recede from each other due to the expansion on the cosmological scale as discussed in this paper.

The problem of the origin of the cosmic acceleration and dark energy will here be studied – from first principles and in an assumed flat universe – by requiring only that the exact transport equation (the Boltzmann transport equation) for quanta propagating through space should be Lorentz-covariant and dispersion-free, and equivalent to a wave equation as given, *e.g.*, in electromagnetism.

It will be shown that this simple and natural assumption leads to a condition which an observer would interpret as an accelerated expansion of the universe, and also a condition in which huge amounts of quanta are seemingly released, resembling the multiplication process in a nuclear fission explosive – albeit on a quite different time-scale. It is in this paper suggested that this could be a possible origin of dark energy, and explain the huge amounts of dark energy now present in the universe.

It should be emphasized that the mechanism studied here implies that dark energy and the accelerated expansion of the universe thus are two *independent* consequences of the requirement of Lorentz-covariant quantum transport. Dark energy and the accelerated expansion are hence not dynamically connected to each other – the accelerated expansion is *not* driven by any pressure from dark energy.

The first calculations as outlined above were given in detail in a paper [13] published some thirty-five years ago, but since in particular the predicted quantum multiplication process seemed to have no relationship to the astronomical picture at the time, the paper had very little impact. Now, however, the problem of the origin of dark energy and the observed accelerated expansion of the universe may perhaps make it meaningful to carry the ideas in the 1975 paper to their logical conclusion. In order to give the basis for the discussion later in this paper, the 1975 paper will here now first be recapitulated (although the reader is encouraged to consult the original paper for the somewhat more detailed derivation given there).

## 2. Boltzmann's transport equation

The time-dependent propagation of neutral quanta (such as, *e.g.*, in gamma radiation) moving with the velocity of light $c$ through a medium, with which they interact by localized collisions, is rigorously described by the Boltzmann transport equation [14],

$$\frac{\partial f(\mathbf{r}, t, \mathbf{\Omega})}{c\, \partial t} = -\, \mathbf{\Omega} \cdot \nabla f(\mathbf{r}, t, \mathbf{\Omega})$$
$$+ \int \Sigma(\mathbf{r}, t, \mathbf{\Omega}')\, K(\mathbf{r}, t, \mathbf{\Omega}' \to \mathbf{\Omega})\, f(\mathbf{r}, t, \mathbf{\Omega}')\, d\mathbf{\Omega}'$$
$$-\Sigma(\mathbf{r}, t, \mathbf{\Omega}) f(\mathbf{r}, t, \mathbf{\Omega}) + S(\mathbf{r}, t, \mathbf{\Omega}), \qquad (1)$$

where $S(\mathbf{r}, t, \mathbf{\Omega})$ is a source term, and $f(\mathbf{r}, t, \mathbf{\Omega})$ is the angular flux in direction $\mathbf{\Omega} = (\Omega_x, \Omega_y, \Omega_z)$ at point $\mathbf{r} = (x, y, z)$ and time $t$. Possible interactions with the medium through which the quanta propagate are described by the interaction cross section $\Sigma(\mathbf{r}, t, \mathbf{\Omega})$, and where the kernel $K(\mathbf{r}, t, \mathbf{\Omega}' \to \mathbf{\Omega})$ then describes how quanta may become scattered from direction $\mathbf{\Omega}'$ to direction $\mathbf{\Omega}$, and/or partially absorbed or multiplied (like neutrons in fission) in the process. It should be emphasized that the Boltzmann transport equation is a rigorous continuity equation for the angular flux, and is exact as long as the angular flux is sufficiently low so that the effects of particle-particle interactions between the propagating quanta themselves can be neglected.

In comparison, the diffusion equation is an approximate equation for the total flux $\Phi(\mathbf{r}, t)$, defined as

$$\Phi(\mathbf{r}, t) = \int f(\mathbf{r}, t, \mathbf{\Omega})\, d\mathbf{\Omega}. \qquad (2)$$

The diffusion equation can be derived from the Boltzmann transport equation (1) above, but with necessity contains serious approximations, mainly because it does not involve the angular distribution of the flux, and the diffusion equation also describes an infinite propagation velocity. Nevertheless, for quanta undergoing isotropic scattering in a homogeneous medium, the quantum propagation as described by the flux $\Phi(\mathbf{r}, t)$ in (2) can be derived [13] rigorously from the Boltzmann transport equation (1) to take the form of the "telegrapher's equation" [15],

$$\Delta\Phi - \frac{\partial^2\Phi}{c^2 \partial t^2} - \left(\frac{1}{3D} + \Sigma_a\right)\frac{\partial\Phi}{c\partial t} - \left(\frac{\Sigma_a}{3D} + \frac{\partial\Sigma_a}{c\partial t}\right)\Phi$$
$$+ \left(\frac{1}{3D} + \frac{\partial}{c\partial t}\right)S = 0. \qquad (3)$$

where

$$D = \frac{1}{3\Sigma} \qquad (4)$$

$$\Sigma_a = (1 - \nu)\Sigma \qquad (5)$$

with

$$\nu = \int K(\mathbf{r}, t, \mathbf{\Omega}' \to \mathbf{\Omega})\, d\mathbf{\Omega}', \qquad (6)$$

and where, *e.g.*, the value $\nu = 0$ corresponds to pure absorption, $\nu = 1$ to pure scattering, and $\nu > 1$ to a multiplying medium. The condition of isotropic scattering in a homogeneous medium implies $\Sigma(\mathbf{r}, t, \mathbf{\Omega}) = \Sigma(t)$, so that $\Sigma_a = \Sigma_a(t)$.

## 3. Special case – wave equation

Due to the third and fourth terms on the left-hand side of the telegrapher's equation (3) above, this equation will not be Lorentz-covariant and it will also display dispersion – two properties that would make it incompatible with a wave equation derived from, *e.g.*, electromagnetism. However, we note that the telegrapher's equation above is compatible with a Lorentz-covariant and dispersion-free quantum propagation as described by a wave equation if the third and fourth terms in (3) satisfy the following two conditions,

$$\frac{1}{3D} + \Sigma_a = 0, \tag{7}$$

$$\frac{\Sigma_a}{3D} + \frac{\partial \Sigma_a}{c \partial t} = 0, \tag{8}$$

*i.e.*

$$\frac{\partial \Sigma_a}{c \partial t} - \Sigma_a{}^2 = 0, \tag{9}$$

which has the solution ($R$ is an integration constant)

$$\Sigma_a = \frac{-1}{R + ct}. \tag{10}$$

From (4) and (7) we see that $\Sigma_a = -\Sigma$, and from (10) and (5), respectively, we thus get

$$\Sigma = \frac{1}{R + ct} \tag{11}$$

$$\nu = 2. \tag{12}$$

Note that the second order time derivative in the basic wave equation (3) [after setting (7) and (8)] may lead to a wavelength with an arbitrarily much shorter characteristic length than the parameter $R$ in (10). This wavelength corresponding to the time derivative may easily be in, *e.g.*, the optical region, despite the fact that the parameter $R$ may possibly be up to the order of the extension of the observable universe.

## 4. The Pareto distribution

The Lorentz-covariant and dispersion-free quantum transport derived above is thus a Markov process with a multiplication factor of 2, and a cumulative path length distribution function given by the following expression [16],

$$F(s) = 1 - \exp\left[-\int_0^s \frac{ds'}{R + s'}\right] \tag{13}$$

or evaluated

$$F(s) = 1 - \frac{R}{R + s} \tag{14}$$

*i.e.* a Pareto distribution of the second kind [17], a distribution more commonly encountered in economics and sociology. The frequency distribution corresponding to the above cumulative path length distribution in (14) is

$$f(s) = \frac{R}{(R + s)^2} \tag{15}$$

A typical Pareto frequency distribution is illustrated in figures 1 and 2 below.

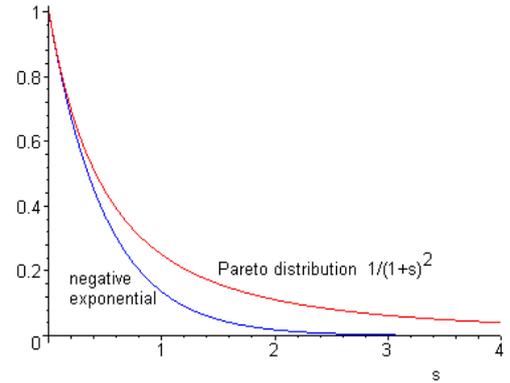

**Figure 1.** Pareto distribution $f(s) = 1/(1 + s)^2$ compared to $f(s) = e^{-2s}$, coinciding for small $s$. Note the frequent occurrence of longer path lengths $s$ in the Pareto distribution (cf figure 2).

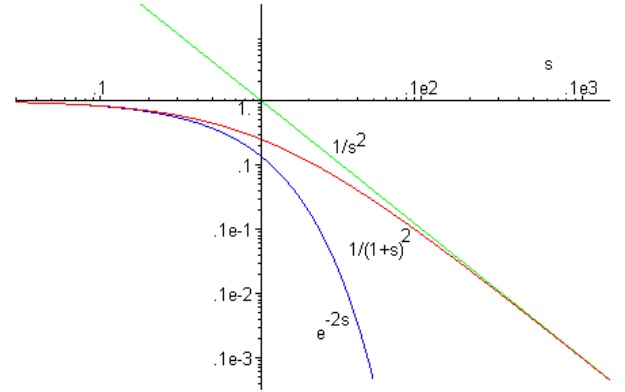

**Figure 2.** The Pareto distribution and negative exponential distribution in figure 1 displayed in a loglog diagram in comparison with an inverse square distribution. We see how the Pareto distribution agrees with a negative exponential for small $s$ and with an inverse square law for large $s$.

The path-length distribution above is thus a rational function in contrast to the exponential path-length distribution normally encountered in transport theory. In particular – and quite different from normal particle propagation in a medium – the mean free path corresponding to the Pareto distribution above will be infinite,

$$\int_0^\infty s\, f(s)\, ds = \int_0^\infty \frac{s\, R}{(R+s)^2}\, ds = \infty \qquad (16)$$

Figures 3 and 4 below show the result of a computer simulation, in which a particle is started at the origin and then followed through successive collisions as described by the Pareto distribution, and where each collision becomes the starting point for two new trajectories.

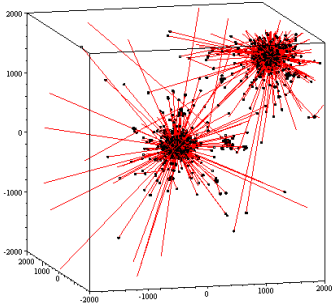

**Figure 3.** Monte-Carlo simulation of the spatial distribution of collision points (black dots) after 131,070 collisions (16 generations of particle doubling) according to figure 1 and starting at (0, 0, 0). We see how collisions tend to lump together in clusters, but also – since the mean-free path in the Pareto scattering process is infinite – may suddenly take long leaps to new collision areas.

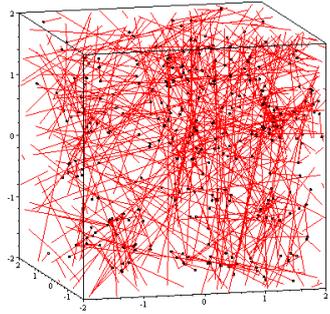

**Figure 4.** Detail of figure 3 around the origin. We see how collisions appear against a substantial, essentially homogeneous background of trajectories from collisions far away. This substantial background of quantum trajectories is here postulated to constitute dark energy.

As a prelude to the discussion below in connection with figures 6 – 8, the figure 5 below shows the radial distribution of particles as function of time from a Monte-Carlo simulation of a Pareto transport and particle doubling as discussed above, here with 200 particles started at radius $r = 0$ and time $t = 0$.

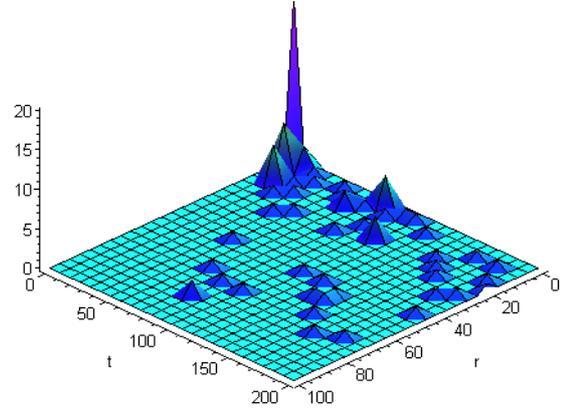

**Figure 5.** Monte-Carlo simulation of the number of particles as function of radius $r$ and time $t$ for transport with the Pareto distribution in (14) and 6 generations of particle doubling as in (12), and with 200 particles started at $r = 0$, $t = 0$, and with $c = 1$. There is a hint of a wave front $r = t$ followed by random collisions (cf figure 6).

## 5. Exponentially accelerated expansion

The somewhat peculiar transport process described above with time-varying cross-sections and quantum creation will now be further analysed by demonstrating how it can be considered as a particular representation of a much simpler transport process.

The simplest nontrivial transport is linear transport in an infinite, homogeneous and isotropic medium with pure isotropic scattering described by a scattering cross-section $\Sigma_0$, constant in space and time, and a multiplication factor $\nu_0 = 1$ (*i.e.* with kernel $K = 1/4\pi$ above). As discussed above, the quantum transport will then be exactly described by the following Boltzmann equation for the angular flux $\varphi(\boldsymbol{\rho}, \tau, \boldsymbol{\omega})$ in space coordinates $\boldsymbol{\rho}$, time $\tau$, and direction $\boldsymbol{\omega}$,

$$\frac{\partial \varphi(\boldsymbol{\rho}, \tau, \boldsymbol{\omega})}{c\, \partial \tau} = -\,\boldsymbol{\omega} \cdot \nabla \varphi(\boldsymbol{\rho}, \tau, \boldsymbol{\omega}) + \int \frac{\Sigma_0}{4\pi}\, \varphi(\boldsymbol{\rho}, \tau, \boldsymbol{\omega})\, \mathrm{d}\boldsymbol{\omega}$$
$$- \Sigma_0\, \varphi(\boldsymbol{\rho}, \tau, \boldsymbol{\omega}) + S(\boldsymbol{\rho}, \tau, \boldsymbol{\omega}), \qquad (17)$$

which as discussed above will not lead to a Lorentz-covariant and dispersion-free wave equation in the world $\boldsymbol{\rho\tau}$.

However, we may transform the above Boltzmann equation (17) into one which does represent a Lorentz-covariant and dispersion-free transport by making the following variable transformations,

$$d\boldsymbol{\rho} = \alpha\, d\boldsymbol{r} \qquad (18)$$

$$d\tau = \alpha\, dt \qquad (19)$$

$$\varphi(\boldsymbol{\rho}, \tau, \boldsymbol{\omega}) = \alpha \, f(\boldsymbol{r}, t, \boldsymbol{\Omega}) \qquad (20)$$

where

$$\alpha = \frac{2}{\Sigma_0}(R + ct)^{-1}. \qquad (21)$$

Making these transformations in the simple, non-Lorentz-covariant transport equation (17) above, we get

$$\frac{\partial f(\boldsymbol{r}, t, \boldsymbol{\Omega})}{c \, \partial t} + f(\boldsymbol{r}, t, \boldsymbol{\Omega}) \frac{\partial \alpha}{c \, \alpha \, \partial t} = -\boldsymbol{\Omega} \cdot \nabla f(\boldsymbol{r}, t, \boldsymbol{\Omega})$$
$$+ \int \frac{1}{R + ct} \frac{2}{4\pi} f(\boldsymbol{r}, t, \boldsymbol{\Omega}) \, \mathrm{d}\boldsymbol{\Omega} - \frac{2}{R + ct} f(\boldsymbol{r}, t, \boldsymbol{\Omega}), \quad (22)$$

which since $\partial \alpha / \partial t = -c \, \alpha / (R + ct)$ simplifies to

$$\frac{\partial f(\boldsymbol{r}, t, \boldsymbol{\Omega})}{c \, \partial t} = -\boldsymbol{\Omega} \cdot \nabla f(\boldsymbol{r}, t, \boldsymbol{\Omega})$$
$$+ \int \frac{1}{R + ct} \frac{2}{4\pi} f(\boldsymbol{r}, t, \boldsymbol{\Omega}) \, \mathrm{d}\boldsymbol{\Omega} - \frac{1}{R + ct} f(\boldsymbol{r}, t, \boldsymbol{\Omega}), \quad (23)$$

and which we can identify from (1) and (6) to have $\Sigma = 1/(R + ct)$ and $\nu = 2$, and hence corresponds to a Lorentz-covariant and dispersion-free transport as discussed at (11) and (12) above.

We note from the two transformations $d\boldsymbol{\rho} = \alpha \, d\boldsymbol{r}$ and $d\tau = \alpha \, dt$ in (18) and (19) above that the velocity of light is equal to $c$ in both the $\boldsymbol{\rho}\tau$ and $\boldsymbol{r}t$ systems, and that by using (21) the relationship between $\tau$ and $t$ can be derived from the following equation

$$d\tau = \alpha \, dt = \frac{2 \, dt}{\Sigma_0 \, (R + ct)} \qquad (24)$$

which integrates to

$$\tau = \frac{2}{c \, \Sigma_0} \, ln\left(1 + \frac{ct}{R}\right), \qquad (25)$$

or in units so that $2/(c \, \Sigma_0) = 1$ and $c/R = 1$,

$$\tau = ln(1 + t), \qquad (26)$$

*i.e.*

$$t = e^\tau - 1. \qquad (27)$$

Differentiating (27), *i.e.*

$$dt = e^\tau d\tau, \qquad (28)$$

and comparing with the transformation $d\tau = \alpha \, dt$ in (19), we thus get $\alpha = e^{-\tau}$, and hence $d\boldsymbol{\rho} = \alpha \, d\boldsymbol{r} = e^{-\tau} d\boldsymbol{r}$, *i.e.*

$$d\boldsymbol{r} = e^\tau \, d\boldsymbol{\rho} \, . \qquad (29)$$

Integrating (29) (with suitable origins) we thus get

$$\boldsymbol{r} = e^\tau \boldsymbol{\rho}. \qquad (30)$$

As defined above, the $\boldsymbol{\rho}\tau$ system is a system with "classical" spacetime. Compared to the simple, classical transport in the $\boldsymbol{\rho}\tau$ system, the Lorentz-covariant system $\boldsymbol{r}t$, with which the $\boldsymbol{\rho}\tau$ system coincides for $t = \tau = 0$, is thus subjected to an exponentially accelerated expansion as given by (29) and (30), showing how a line element $d\boldsymbol{r}$ in the Lorentz-covariant $\boldsymbol{r}t$ system increases exponentially with time $\tau$ [and correspondingly for a time element $dt$ as shown in (28)].

## 6. Discussion of the $\boldsymbol{\rho}\tau$ and $\boldsymbol{r}t$ systems

It is illustrative to consider Monte-Carlo simulations of the particle transport in the $\boldsymbol{\rho}\tau$ and $\boldsymbol{r}t$ systems discussed above. First of all, it is then necessary to consider what happens to figure 5 with better statistics, *i.e.* with more particles started at $r = 0$ and $t = 0$ in the $\boldsymbol{r}t$ system, as is shown in figure 6 below.

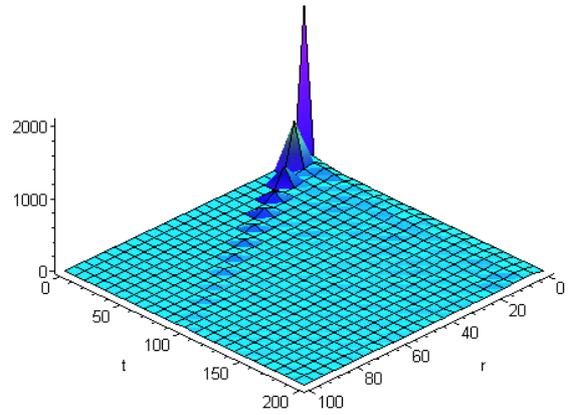

**Figure 6.** Monte-Carlo simulation as in figure 5, but with 20,000 particles started. The wave front $r = t$ is now the dominating feature.

We see that particles then accumulate in the wave front, whereas behind it the particle distribution becomes smeared out to an essentially homogeneous background. This thus illustrates how the combination of the Pareto distribution in (14) and the particle doubling in (12) indeed leads to what looks like propagation of a wave front, despite the fact that it is very much the result of a process of particle scattering.

It will now be illustrated how this wave-type transport in the $\boldsymbol{r}t$ system can be regarded as the result of a transformation as in (18) − (20) of the simpler transport in the $\boldsymbol{\rho}\tau$ system mentioned above, *i.e.* as a transformation of a pure isotropic scattering with multiplication factor $\nu = 1$ in an infinite, homogeneous and isotropic medium. In the $\boldsymbol{\rho}\tau$ system we then have a particle distribution as in figure 7 below as function of radius $\rho$ and time $\tau$, where a Monte-Carlo simulation gives a slowly expanding blob of collision points and connecting trajectories.

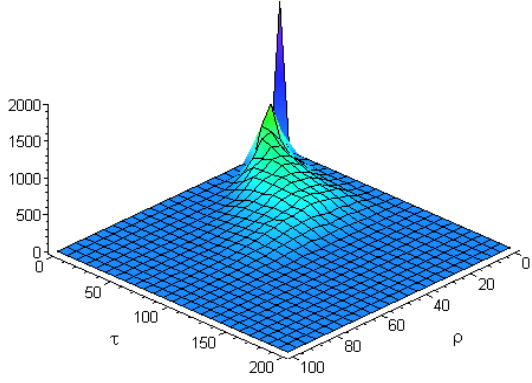

**Figure 7.** Monte-Carlo simulation with 15,000 particles started at $\rho = 0$, $\tau = 0$, and with $c = 1$, and followed for 50 collisions in the simple $\boldsymbol{\rho\tau}$ system: an infinite, homogeneous and isotropic medium with pure isotropic scattering and a multiplication factor $\nu = 1$.

Transforming the $\rho\tau$ system in figure 7 to the $\boldsymbol{rt}$ system according to (27) and (30), we then get a picture as in figure 8 below, which apart from minor statistical scatter well agrees with figure 6 above, thus giving a numerical illustration of the transformation defined by (27) and (30) as discussed in the previous Section.

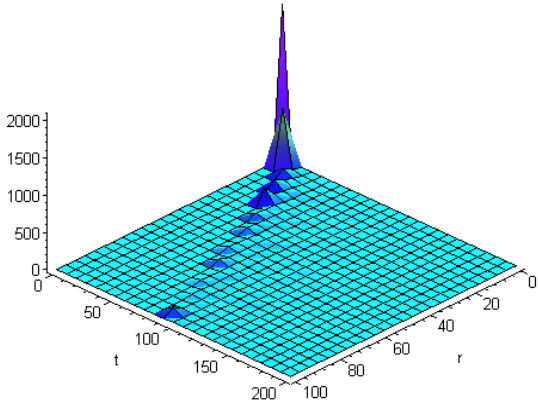

**Figure 8.** Transformation according to (27) and (30) of the simple transport in the $\boldsymbol{\rho\tau}$ system in figure 7, giving a result essentially as in figure 6.

## 7. Quantum multiplication *vs* expansion

It might perhaps superficially look as if the exponentially accelerated expansion derived above would be a dynamical effect of the multiplication of the quantum flux described earlier in this paper. However, it should be emphasised that this flux multiplication and the accelerated expansion are here completely unrelated phenomena dynamically. In this case, they are just two essentially independent consequences of the change of metric required by Lorentz-covariance of the transport equation as discussed above.

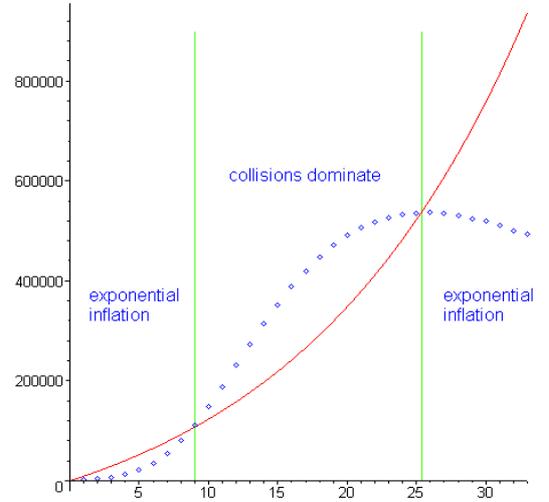

**Figure 9.** Monte-Carlo simulation (dots) of the number of collisions as function of time from a Pareto distribution [1000 particles started at time zero, 15 generations of particle doubling according to (12)]. The solid curve is an example of a corresponding, exponentially accelerated expansion as function of time as given by (30). In this example, there is an initial period when the exponentially accelerated inflation dominates, which is then replaced by an epoch when the collisions dominate, then being replaced again by an epoch when the exponentially accelerated expansion dominates.

The flux multiplication discussed above in Sect. 3, and the exponentially accelerated expansion discussed in Sect. 5, will in general display different behaviour as functions of time as is illustrated by the example in figure 9 above, which gives the results of a Monte-Carlo simulation based on the above derivation. The example in figure 9 shows a short initial period when exponentially accelerated inflation dominates the evolution of the universe. This short period of initial exponential inflation is then followed by a long period when the exponential expansion is hidden behind a collision-dominated world, then again followed by a recent period dominated by the exponential acceleration.

Thus the general behaviour in this simulated example is in crude qualitative agreement with the current cosmological picture. However, it should be emphasised that the details of the simulation in figure 9 are dependent, *e.g.*, on the assumed number of generations and on the exponential parameter used in the accelerated expansion. As will be further discussed below, it is in particular possible that the late exponential expansion $e^{\tau} = 2^{\tau/\ln(2)}$ given in (30) above could in the present epoch more or less exactly match the quantum duplication function of type $2^t$ as given by (12), and together give an essentially constant mass-energy in the universe as function of time.

## 8. Is energy conserved?

The mechanism described above to ensure that particle transport in the $rt$ system is relativistically covariant thus requires an accelerated expansion and quantum duplication to be forced upon the system. This raises a serious question on the conservation of mass-energy in the $rt$ system in this process. This question will now be addressed.

Noether's symmetry theorem [18], which can be shown to be valid under very general conditions, states that symmetry properties of a system lead to conservation laws: symmetry under translation corresponds to conservation of momentum, symmetry under rotation corresponds to conservation of angular momentum, symmetry in time corresponds to conservation of energy, etc.

Although thus conservation of energy is normally ascertained by symmetry with respect to time, such symmetry may be in doubt on cosmological time scales and for objects of galactic dimensions subjected to accelerated expansion. Especially for phenomena like the combined accelerated expansion and particle duplication described in this paper, there seems to be no guarantee that the first law of thermodynamics in its normal formulation should any longer be strictly valid for the system.

On the other hand, above we also considered the system $\rho\tau$ with an assumed infinite, homogeneous and isotropic medium with pure isotropic scattering, and a constant scattering probability in space and time. Such a system will be symmetric in time also over cosmological time scales and galactic dimensions, and the validity of the first law of thermodynamics will thus be ascertained in the $\rho\tau$ system.

The relationship between the $\rho\tau$ system and the $rt$ system as given in (27) and (30) above will then also put stringent conditions on the energy content of the $rt$ system as function of time, even though energy may not be strictly conserved in the $rt$ system over cosmological times. Note also the possibility discussed in the previous section that the exponential expansion and quantum duplication in the $rt$ system may match each other to produce a constant average mass-energy density in the universe as function of time despite its expansion.

## 9. Paradigm shift?

What is left out of the discussion in this paper is the nature of the "classical, non-Hubble world" $\rho\tau$ discussed above, which in principle could be an essentially eternal, steady-state world, and compared to which we see distant galaxies in the Lorentz-covariant world $rt$ like in perspective distortion due to the requirement of Lorentz-covariance.

More critically, what is also left out of the discussion above is the nature of the quanta assumed to constitute dark energy, and the related question of the physical mechanism behind the duplication process by which these quanta get multiplied and so pervade the universe. This latter question will now be briefly addressed.

The theory of relativity has led to a paradigm shift: we no longer question what detailed dynamical mechanisms cause the tension in fast-moving objects to make them shrink in the direction of motion, or influence the working of the balance-wheel in clocks to make them go more slowly at very high speed; we know that these phenomena are due to basic properties of spacetime, not to any particular mechanical effects. After over ten years of fruitless struggle to understand dark energy, maybe we now have to accept that dark energy, and its relationship to the accelerated expansion of the universe, similarly cannot be described in mechanical terms, but are still other observational effects of the relativistic properties of spacetime as discussed in this paper.

## 10. Summary

By assuming only that the time-dependent deep-space propagation of quanta is governed by a rigorous, Lorentz-covariant continuity equation, the following observational characteristics of the universe can be deduced:

1. The universe is subjected to an apparent exponentially accelerated expansion as given in (29) and (30).
2. The quantum propagation from distant objects is subjected to an apparent duplication process as given in (12), and assumed to be the source of dark energy. As a result, the dark energy part of the total mass-energy content of the universe increases as $2^t$ with time $t$ (suitably scaled).
3. After sufficiently long time, dark energy may thus constitute a dominating part of the mass-energy content of the universe.
4. The accelerated expansion, and the amount of dark energy created, are two independent consequences of the Lorentz-covariant transport and are not dynamically connected to each other.
5. The quantum duplication process producing dark energy may possibly more or less balance the exponentially accelerated expansion to give an essentially constant average mass-energy density in the universe as function of time despite the expansion.

## Acknowledgments

The author is indebted to the late Dr Staffan Söderberg for valuable mathematical assistance in the initial phase of this project. My sincere thanks also go to the editors at Physica Scripta for their professionalism and to the referee for a great number of relevant and valuable questions and comments. I also wish to thank the many readers who have critically studied the successive updates of my manuscript arXiv:1003.3870 [astro-ph.CO] on http://arXiv.org/ and asked all the relevant questions that are now summarized in the collection of FAQ also filed there, with special thanks to Dr Hans-Olov Zetterström for many fruitful discussions and a great number of relevant and valuable questions and suggestions. I am also happy to express my gratitude to Dr Andrea Di Vita for important questions (e.g., FAQ #4a) on assumptions made in my derivation.

# $F$requently $A$sked $Q$uestions

(including some questions that should have been asked, but haven't)

**on "Lorentz-covariant quantum transport and the origin of dark energy"**
**(arXiv:1003.3870) by Arne Bergstrom**

---

## FAQ #1 (page 1, column 2)

"You say that available WMAP and other data indicate that the universe is flat on cosmological scales, i e that gravitational effects can be neglected on such scales, as is assumed in your paper. I find it hard to agree with that. The only force on cosmological large scales maybe gravity, so then how can we neglect gravitational effects?"

ANSWER: In my paper I state that recent observations (my ref [11]) show that the universe is flat on cosmological scales (i e beyond the scale of galaxies). This thus means that on these scales it is described by special relativity. The only force on cosmological scales may be gravity, as you say, but on such scales there is in addition a requirement due to special relativity for the quantum transport to be Lorentz-covariant. This requirement forces the universe into an apparent accelerated expansion (and quantum duplication) as I derive in my paper. This forced expansion then completely dominates over any gravitational attraction on this scale – as is astronomically observed.

---

## FAQ #2 (page 2, column 1)

"In the standard cosmological models we argue that dark energy may be the cosmological constant (vacuum energy), scalar field or others. What are the quanta you assume to constitute dark energy? And what are the properties of these quanta?"

ANSWER: No one seems to know what quanta are involved in dark energy. However, the purpose of my article is to show that the very basic and natural assumption of Lorentz covariance of the general transport of quanta – of any kind - in flat spacetime indeed leads to an expanding metric. Basically, the derivation is thus not dependent on what type of quanta is assumed to be involved. The equation I use to describe the time-dependent propagation of these quanta in the paper is an equation used since more than fifty years to describe, e g, the propagation of gamma rays or neutrons in a medium, and it is exact for all types of neutral quanta interacting by localised collisions with a medium. The purported value and novelty of the approach in my paper is that it thus shows that one simple, very fundamental mechanism may lie behind both the observed exponential expansion of the universe and – independently - the existence of huge amounts of dark energy.

---

## FAQ #3 (page 2, column 2)

"In Sect 2 you state that the telegrapher's equation is valid 'for quanta undergoing isotropic scattering in a homogeneous medium'. Does this mean that it is exact for an arbitrary anisotropic flux, e.g., close to a point source in a homogeneous medium?"

ANSWER: No, it is exact only for isotropic scattering in an infinite homogeneous medium, *i.e.* not close to a solitary point source or close to a boundary to a different medium. "Close" is then to be understood as within several mean free paths, outside which the particle propagation is expected to have averaged out to something comparatively isotropic, and where the telegrapher's equation would be valid.

---

## FAQ #4 (page 3, column 1)

"In your first description of the quantum propagation on pages 2 and 3, you get a wave moving with velocity *c*. This means an expansion, but not an accelerating one, does it not"

ANSWER: We need to distinguish between the particle propagation and the motion of the medium. The propagation velocity of the wave on pages 2 and 3 is *c* as you say. But this is the velocity of the quanta and has nothing to do with the motion of the medium through which the quanta propagate. Irrespective of the motion of the quanta, the medium can be stationary, expanding, or subject to an accelerating expansion, just like in a nuclear reactor (or nuclear explosive) the gamma rays move with the velocity of light in a stationary (or slowly moving) medium. The motion of the medium is described by the time-dependent macroscopic cross-section $\Sigma(t)$ as given at the bottom of page 2.

---

## FAQ #4a (page 3, column 1)

"In your equation (3) you set the third and fourth terms separately to zero. But should they not instead be set to zero together? Instead of your (7) and (8), shouldn't we have

$$\left(\frac{1}{3D} + \Sigma_a\right)\frac{\partial \Phi}{c\partial t} + \left(\frac{\Sigma_a}{3D} + \frac{\partial \Sigma_a}{c\partial t}\right)\Phi = 0 \qquad \text{(i)}$$

which is obviously a more general relationship than your (7) and (8)?"

ANSWER: Please remember that we are considering linear quantum transport. Then, as I say in Sect 2 in my paper, the angular flux is assumed to be "sufficiently low so that the effects of particle-particle interactions between the propagating quanta themselves can be neglected". To avoid the type of concern you express, I should have added "... and that the particles also do not influence the medium through which they propagate". These are the basic assumptions made when deriving the Boltzmann transport equation (1). Thus we need to consider the material characteristics $D$, $\Sigma_a$, $\nu$, etc as independent of the flux $\phi$. In a nuclear reactor, for instance, the Boltzmann equation can be used to calculate the neutron flux at a specific time, but then new transport parameters $D$, $\Sigma_a$, $\nu$, etc may have to be introduced as the fuel gets depleted.

The "vacuum quantities" $D$, $\Sigma_a$, $\nu$, etc in our case may vary with time but they are independent of the flux $\phi$. They are there all the time, flux or no flux. They may vary so that the flux is Lorentz covariant, but they will do so in the same way even for an infinitesimal flux. The flux $\phi = \phi(\mathbf{r}, t)$, on the other hand, will vary depending on the source term $S = S(r, t)$ in (3). Thus, as I see it, it makes no sense in this case to let the quantities $D$, $\Sigma_a$, $\nu$, etc vary with the flux $\phi$ or its time derivative in order to satisfy (i). We should set both terms to zero independently as in my paper.

---

## FAQ #5 (page 3, column 1)

"When you choose a $\Sigma_a$ according to (10), you seem to introduce a universal time. Is that not a problem? I mean, you get a Lorentz-covariant equation − but at the expense of the requirement of a universal time."

ANSWER: No, any universal time is not required. We can always rewrite the denominator in (10) as $R + ct = R_0 + c(t + t_0)$, where we can choose $t_0$ as we like (although $R_0$ then of course also changes).

---

## FAQ #6 (page 3, column 1)

"I find it difficult to understand the quantum duplication in (12), which together with the exponential expansion is forced upon the $rt$ world compared to the $\rho\tau$ world by the requirement of Lorentz-covariance. Can one somehow visualize this duplication? What is conserved in the duplication process?"

ANSWER: There is a kind of vanishing puzzle with, say, eight men. You turn the centre part − and then you have suddenly somehow mysteriously got nine men instead! But where did the extra man come from? (http://www.samloyd.com/vanishing-puzzles/index.html). The quantum duplication works somewhat similarly: We have an exponential expansion of space with time according to (30). The quantum duplication $2^t = e^{t \, ln2}$ also means an exponential increase with time, and can then keep the quantum density constant despite the expansion. As I try to say in Sect 8, it may not be necessarily relevant to discuss the detailed dynamics of the quantum multiplication, just as it is not relevant to discuss in mechanistic terms how the Lorentz contraction of an object is caused.

---

## FAQ #7 (page 3, column 1)

"The quanta you discuss are subjected to a duplication process. But must there not be also 'normal' quanta in the universe that do not experience such duplication? If so, what then differs those normal quanta from those that get duplicated?"

ANSWER: In principle nothing. Normally, all quanta interact in the usual way with the matter, or the remnants of matter, through which they pass. The only exceptional situation is at extreme distances far outside their parent galaxy, or clusters of galaxies, because in that extremely tenuous environment their propagation still needs to be governed by a wave equation (which is Lorentz-covariant and dispersion-free). But the propagation of the quanta in this tenuous environment is rigorously described by the Boltzmann transport equation, which is *not* Lorentz-covariant and dispersion-free. This circle can only be squared by imposing the conditions derived in Sect 5, which thus correspond to an exponentially accelerated expansion of the universe at cosmological scales, and the strange quantum multiplication described in the paper − both effects in agreement with astronomical observations.

---

## FAQ #7a (page 3, column 1)

**"You start with Boltzmann's transport equation (1) in your paper. This equation definitely describes an irreversible process. Then you make variable transformations so that you get a Lorentz-covariant and dispersion-free wave equation. Since it is a wave equation, it thus describes a reversible process. But how can we make a variable transformation that transforms an equation that describes an irreversible process into one that describes a reversible process? Isn't that what you need a Maxwell's demon for?"**

<u>ANSWER:</u> Maybe this says something deep about the cosmological expansion. Generally speaking, irreversible processes raise entropy. It is likely that the requirement of Lorenz' covariance of the Boltzmann equation puts constraints on the evolution of the entropy of the Universe during the cosmological expansion.

## FAQ #8 (page 3, column 1)

**"In the last paragraph of Sect 3, you talk about arbitrarily small wavelengths in, e.g., the optical region. Is that something that can be observed?"**

<u>ANSWER:</u> What I am trying to say here is that the telegrapher's equation is valid not just for characteristic lengths corresponding to the extension of the universe, but also for much smaller characteristic lengths, e.g., as in ordinary optical wave propagation in a normal medium. The reason for this comment is to allow for the possibility that the quanta involved in the mechanism discussed in my paper could conceivably be ordinary photons.

## FAQ #9 (page 5, column 1)

**"I think it is unclear that Eq (29) and Eq (30) give the solution of an accelerating expansion of the universe. Which is the scale factor of our universe, is it $r$ or $\rho$? Is it that $\frac{d^2 r}{dt^2} > 0$ or $\frac{d^2 \rho}{d\tau^2} > 0$ mean an accelerating expansion of the universe?"**

<u>ANSWER:</u> The "scale factor of our universe" is $r$ in my paper. In contrast, the "scale factor" $\rho$ defined in Sect 5 in my paper corresponds to a classical Newtonian universe with gravitation of the type that describes the motion of planets in the solar system (but not the universe on cosmological scales or close to black holes). Thus the world $\rho\tau$ is a classical Newtonian universe with no cosmological expansion. Equations (27) and (30) then describe the relationship between this "classical" world $\rho\tau$ and the "real" Lorentz-covariant world $rt$. From (27) and (30) we see that the Lorentz-covariant world $rt$ corresponds to a world in exponentially accelerating expansion on cosmological scales compared to the classical Newtonian world $\rho\tau$ – as is astronomically observed.

## FAQ #10 (page 5, column 2)

**"In Figure 6 you see a distribution in the form of 'islands' along the line $r = t$. Does this have any significance?"**

ANSWER:  No, this is just a numerical effect of the rectangular grid. In some places we happen to get a registration of all the wave front in one grid point only, whereas in other places the wave front gets distributed over a number of adjacent grid points and then shows up as a much lower value in all of them.

---

## FAQ #11 (page 5, column 2)

"I have some difficulties in understanding whether we live in the $rt$ world or in the $\rho\tau$ world."

ANSWER:  The simplest answer is perhaps: "In both". With regard to phenomena and observations on smaller than extragalactic scales – i.e. almost everywhere under normal circumstances - the $rt$ world and the $\rho\tau$ world are identical. But on extragalactic scales the requirement of Lorenz-covariant quantum transport forces the exponential expansion and quantum duplication to appear – this is thus the exclusive realm of the $rt$ world.

---

## FAQ #12 (page 6, column 2)

"It seems difficult reconcile your explanation of the origin of dark energy to current cosmological observations. For example, we know that our universe has two accelerating expansion epochs. One happened when the universe was very young, about $10^{-35}$ s old, what we now call inflation. Another is the late-time accelerating expansion, namely the dark energy problem we are discussing. Furthermore, we know that the second accelerating expansion began recently (at redshift z<0.7). So what is the role of the quanta in this paper - inflation or dark energy? If it is the dark energy, how to make the universe begin an accelerating expansion recently?"

ANSWER:  The main role discussed for the quanta in my paper is in dark energy. In principle, the exponential expansion I derive is there all the time, but the conditions around the initial inflation make its application to that period questionable. In broad strokes, however, the simulation in my fig. 9 does show one initial and one late period of exponentially accelerated expansion, and separated by an epoch when the exponential expansion is hidden behind a collision-dominated period. I discuss this point in Sect 7. (Please note that the parameters in my simulation in fig. 9 are chosen to give an illustrative result; as you correctly point out, the initial inflation is of course in reality very much shorter.)

---

4 May 2010

Arne Bergstrom